\documentclass[preprint,aps,showpacs,nofootinbib]{revtex4}

\usepackage{epsfig,amssymb,amsmath}

\def\bea{\begin{eqnarray}}
\def\eea{\end{eqnarray}}
\def\beq{\begin{equation}}
\def\eeq{\end{equation}}

\begin{document}
\draft
\tighten
\preprint{TU-924}
\title{\large \bf
~\vskip 0.4cm
Higgs mixing and diphoton rate enhancement in NMSSM models
}
\author{
Kiwoon Choi$^{a,}$\footnote{email: kchoi@kaist.ac.kr},
Sang Hui Im$^{a,}$\footnote{email: shim@muon.kaist.ac.kr},
Kwang Sik Jeong$^{b,}$\footnote{email: ksjeong@tuhep.phys.tohoku.ac.jp},
Masahiro Yamaguchi$^{b,}$\footnote{email: yama@tuhep.phys.tohoku.ac.jp}
}
\affiliation{
$^a$Department of Physics, KAIST, Daejeon 305-701, Korea \\
$^b$Department of Physics, Tohoku University, Sendai 980-8578, Japan}

\vspace{2cm}

\begin{abstract}

We study the implications of the LHC Higgs signals on the Higgs mixing in
the next-to-minimal supersymmetric standard model (NMSSM).
The Higgs couplings can depart from their values in the standard model (SM)
due to mixing effects.
However the Higgs signal rate in the $WW$ and $ZZ$ channels can remain
close to the SM values, as observed at the LHC, even if the SM-like Higgs boson
with a mass near 125 GeV has a large singlet component.
This allows to get a sizable enhancement in the Higgs to diphoton rate
through the charged-higgsino loop contribution, as well as a sizable reduction
of the Higgs to $b\bar b$ and $\tau\tau$ rates through the mixing effects,
with little deviation in the $WW$ and $ZZ$ signal rates from
the SM prediction.
We find that an enhancement of diphoton signals by a factor of 1.5 or more,
and also a reduction of $b\bar b$ and $\tau\tau$ signals by a factor
of 0.5, can be obtained in the region of parameter space consistent
with the constraints on the higgsino mass parameter and the singlet
coupling to the Higgs doublets, which determine the Higgs mixing.

\end{abstract}

\pacs{}
\maketitle

\section{Introduction}

The ATLAS and CMS collaborations have announced the discovery of
a Higgs boson near 125 GeV with standard model (SM)-like properties
\cite{ATLAS-Higgs,CMS-Higgs}.
It is thus of importance to study if models with extended Higgs
sector, which is motivated to solve various difficulties of the
SM, can accommodate such a SM-like Higgs boson.
In this paper we consider the next-to-minimal supersymmetric
SM (NMSSM).
The NMSSM maintains most of the nice features of the MSSM while
allowing richer physics in the Higgs and neutralino sectors since
it includes a SM singlet which couples to the Higgs doublets
\cite{Review-NMSSM}.

In the NMSSM, the Higgs boson properties can significantly deviate
from the SM ones due to Higgs mixing effects.
Meanwhile the observed Higgs signal rate in the $WW/ZZ$ channel,
for which experimental uncertainties are small compared to those
in other channels, is almost consistent with the SM predictions.
This however does not necessarily demand small Higgs mixing, and
it turns out that the observed signal rate in this channel
can still be explained even if the SM-like Higgs boson has a large
singlet component.
It is then interesting to examine how the signal rate in other
channels are modified when the Higgs boson mixes with the additional
neutral scalars in such a way that the $WW/ZZ$ rate remains close
to the SM rate.

In particular we pay our attention to the possibility that
the observed enhancement in the diphoton rate relative to the SM value,
though we need more statistics to confirm this observation,
is accommodated in the presence
of Higgs mixing effects.\footnote{
It has been noticed that Higgs mixing effects can increase
the diphoton rate above the SM value in the supersymmetric models
\cite{Hall:2011aa,Ellwanger:2010nf,Ellwanger:2011aa,Kang:2012sy,Barger:2012ky},
for which case the $WW/ZZ$ rate is also enhanced by a similar amount,
assuming that the MSSM superparticles are heavy.
}$^{,}$\footnote{
There have been many works studying how to get a large enhancement
in the diphoton signal,
see e.g. refs.
\cite{Blum:2012ii,Carena:2012xa,Joglekar:2012vc,ArkaniHamed:2012kq,Almeida:2012bq,
Cao:2012fz,Benbrik:2012rm}
for general discussion.
In the supersymmetric SM, charged superparticles can give a sizable
contribution to the Higgs coupling to photons if they are light.
For instance, in refs.
\cite{Carena:2011aa,Carena:2012gp,Buckley:2012em,Giudice:2012pf,Ajaib:2012eb,Sato:2012bf},
light third generation sfermions with large left-right mixing have been considered
to achieve this.
It is also possible to consider the contribution from light charginos or
charged Higgs scalar
\cite{King:2012is,Gabrielli:2012hd,Haba:2012zt,Alves:2012ez,SchmidtHoberg:2012yy}.
Alternatively, one may extend the supersymmetric SM to include extra
charged-particles with light masses \cite{An:2012vp,Delgado:2012sm,Basso:2012tr}.
In this case, if colored particles are added, the Higgs production rate
at the LHC would be affected significantly.
}
Indeed this is achievable through the charged-higgsino loop contribution
if the SM-like Higgs boson has a sizable singlet component.
In such a case, the mixing effects generally reduce the $b\bar b$ and $\tau\tau$
rates below the SM values, because the Higgs couplings to $b\bar b$ and $\tau\tau$
are suppressed more strongly than those to the $W/Z$ boson and
the top quark.
On the other hand, the mixing in the neutral Higgs sector depends on
the higgsino mass parameter $\mu$ and the singlet coupling $\lambda$
to the Higgs doublets.
This implies that the size of mixing effects is subject to the LEP bound
on the chargino mass, and the perturbativity limit on $\lambda$,
as well as to the requirement on the mass of the SM-like Higgs boson.
Thus one should consider these, and also other possible constraints on
the model parameters, in order to see how large enhancement can be obtained
in the diphoton channel through the charged-higgsino contribution.
It turns out that an enhancement of diphoton rate, for instance,
by a factor 1.5 or larger, as well as a reduction of $b\bar b$ and $\tau\tau$
rates by a factor of 0.5, can be achieved while having little deviation in the
$WW/ZZ$ rate from the SM value.

This paper is organized as follows.
In section \ref{sec:Higgs-NMSSM}, we will briefly review how the Higgs boson
couples to the SM particles when it mixes with other neutral scalars,
and then examine the Higgs signal rate.
We will focus on the question which mixing angles are allowed for the
SM-like Higgs boson whose signal rates in the $WW/ZZ$ channel are close
to those predicted by the SM.
In section \ref{sec:higgsino-effect}, we will discuss the possible enhancement
of the diphoton rate through the charged-higgsino loop contribution.
The constraints on the model parameters will be explored in detail in section
\ref{sec:constraints}, where we present some examples to see
the mixing effects on the Higgs signal rate in the viable region of parameter
space.
Section \ref{sec:conclusion} is devoted to the conclusion.

\section{Higgs sector of the NMSSM}
\label{sec:Higgs-NMSSM}

In the supersymmetric SM, the properties of the SM-like Higgs boson
crucially depend on the mixing in the neutral Higgs sector.
So an important question is if it is possible to arrange a SM-like Higgs
boson whose properties are consistent with the observation,
in the presence of mixing effects.
We will see that the Higgs signal rate observed at the LHC can
have interesting implications on the Higgs mixing in the NMSSM.

\subsection{Higgs boson with SM-like properties}

We start by discussing how to describe the deviation of Higgs properties
from the SM behavior.
The interactions of a SM-like Higgs boson $h$ at energy scales around 125 GeV
can be examined using the effective action \cite{Carmi:2012in},
\bea
{\cal L}_{\rm eff} &=& C_V
\frac{\sqrt2m^2_W}{v} h W^+_\mu W^-_\mu
+ C_V \frac{m^2_Z}{\sqrt2 v} h Z_\mu Z_\mu
- C_f \frac{m_\psi}{\sqrt2 v}h \bar \psi \psi
\nonumber \\
&&
+\, C_g \frac{\alpha_s}{12\sqrt2\pi v} h G^a_{\mu\nu} G^a_{\mu\nu}
+ C_\gamma \frac{\alpha}{\sqrt2\pi v} h A_{\mu\nu} A_{\mu\nu}
+ \cdots,
\eea
for $\psi$ denoting the SM fermions, and $v\simeq 174 $ GeV.
Here the custodial symmetry is assumed for the Higgs couplings to $W$ and $Z$
bosons as suggested by the electroweak precision measurements,
and the ellipsis denotes Higgs interactions with non-SM particles.
In the SM, the Higgs boson has $C_V=C_f=1$, while its couplings to gluons and
photons are induced only radiatively with the dominant contribution coming from
the $W$-boson and top quark loops.

The Higgs signal rate in the decay channel, $h\to ii$, relative to
the SM prediction is defined by
\bea
R^{ii}_h =
\frac{\sigma(pp\to h)\,{\rm Br}(h\to ii)}{
\sigma(pp\to h)|_{\rm SM}\,{\rm Br}(h\to ii)|_{\rm SM}},
\eea
where $\sigma(h)$ and ${\rm Br}(h\to ii)$ is the total Higgs production
cross section and the branching ratio into the corresponding decay channel,
respectively.
The production of a SM-like Higgs boson at the LHC is dominated
by the gluon-gluon fusion, which receives the main contribution from
the top quark loop.
Let us assume $C_b=C_\tau$ and $C_t=C_c$, which hold at the tree-level
in the MSSM.
Then, for $m_h=125$ GeV, using the well-known production and decay properties
of the SM Higgs boson having the same mass, one finds the Higgs signal
rate in the $WW/ZZ$ channel to be
\bea
\label{RVV}
R^{VV}_h \simeq
\frac{(0.94 C^2_g + 0.12 C^2_V) C^2_V}{0.64 C^2_b + 0.24 C^2_V + 0.12 C^2_t},
\eea
under the assumption that new Higgs decay mode if kinematically allowed
has a negligibly small decay width.
The Higgs signal rate in other channels read
\bea
\label{Rii}
R^{b\bar b}_h &=& R^{\tau\tau}_h
= \frac{C^2_b}{C^2_V} R^{VV}_h,
\nonumber \\
R^{\gamma\gamma}_h &\simeq&
\frac{1.52 C^2_\gamma}{C^2_V} R^{VV}_h.
\eea
In the SM, the Higgs boson has $R^{ii}_h=1$ by definition.
Here one should note that the effective couplings $C_g$ and $C_\gamma$ receive
only loop contributions in the MSSM.
For the case that colored superparticles have masses well above the weak scale,
which we will assume in this paper, the effective Higgs coupling to gluons
and photons are written as
\bea
C_g &\simeq& 1.03 C_t - 0.06 C_b,
\nonumber \\
C_\gamma &\simeq&
0.23 C_t - 1.04 C_V + \delta C_\gamma|_{\rm SUSY},
\eea
for $h$ having $m_h=125$ GeV.
The loop correction from charged superparticles, $\delta C_\gamma|_{\rm SUSY}$,
becomes important when they are light.

The above shows that the signal rates for the decay processes relevant to Higgs
searches are determined by the effective Higgs couplings $C_V$, $C_t$, $C_b$
and $\delta C_\gamma|_{\rm SUSY}$.
We will examine how $C_V$, $C_t$ and $C_b$ are modified by Higgs mixing
effects in the NMSSM, and then move on to discuss how large correction to
$C_\gamma$ is possible from the loops of charged superparticles.
In particular, we focus on the possibility to get a large enhancement of
the Higgs coupling to photons through the contribution from the
charged-higgsino loop, which requires the charged higgsino to have a relatively light mass
and also the SM-like Higgs boson to have a sizable singlet component.

\subsection{Higgs boson in the NMSSM}

The Higgs boson in the supersymmetric SM can have properties which are
significantly deviated from the SM properties due to Higgs mixing
effects.
In particular, NMSSM models have rich Higgs/neutralino physics as the
Higgs sector is extended to include a SM singlet field $S$.
The singlet couples to the Higgs doublets through,
\bea
W = \lambda S H_u H_d + f(S) + (\mbox{MSSM Yukawa terms}),
\eea
so that a higgsino mass parameter is dynamically generated
after it gets a vacuum expectation value.
The effective Higgs $\mu$ and $B\mu$ terms read
\bea
\mu &\equiv& \lambda \langle S \rangle,
\nonumber \\
B\mu &\equiv& A_\lambda \lambda \langle S \rangle
+ \lambda \langle \partial_S f \rangle,
\eea
where $A_\lambda$ is the soft SUSY breaking trilinear parameter, and we
take a field basis such that a bare $\mu$ term if exists is absorbed
into $S$.
The superpotential $f(S)$ is necessary to avoid a phenomenologically
unacceptable visible axion, and various models are classified
according to its form.
In this paper, we do not specify the exact form of $f(S)$, but
will assume that there is no CP violation in the Higgs sector.
Noting that $H^0_u$ and $H^0_d$ mix with the singlet scalar through
the scalar potential terms,
\bea
V_{\rm mix} \,=\, \lambda^2 |S|^2 \left(|H_u|^2+|H_d|^2\right)
+ \left( A_\lambda\lambda SH_uH_d
+ (\partial_S f)^*\lambda H_uH_d +{\rm h.c.} \right),
\eea
we introduce a mass parameter,
\bea
\Lambda \equiv A_\lambda + \langle \partial^2_S f \rangle,
\eea
for discussion on the Higgs mixing, and will treat $\mu$, $B\mu$ and $\Lambda$
as independent parameters as $f(S)$ is needed in any model for the Higgs
sector not to possess a global symmetry, unless $\lambda$ is extremely small.

The neutral Higgs sector of the NMSSM has three CP-even Higgs bosons:
\bea
\hat h &=& \sqrt2\left(
({\rm Re}H^0_d - v\cos\beta)\cos\beta
+ ({\rm Re}H^0_u - v\sin\beta)\sin\beta \right),
\nonumber \\
\hat H &=& \sqrt2\left(
({\rm Re}H^0_d - v\cos\beta)\sin\beta
-({\rm Re}H^0_u - v\sin\beta)\cos\beta \right),
\nonumber \\
\hat s &=& \sqrt2\left(
{\rm Re}S - \langle S \rangle \right),
\eea
where $\langle H^0_u \rangle = v\sin\beta$ and $\langle H^0_d \rangle = v\cos\beta$.
Among them, $\hat h$ behaves exactly like the SM Higgs boson if it does
not mix with the others.
However, since the mass matrix of the hatted fields is given by
\bea
\hat M^2 =
\left(%
\begin{array}{ccc}
  m^2_{\hat h} & \frac{1}{2}(m^2_Z-\lambda^2 v^2)\sin4\beta &
  \lambda v (2\mu-\Lambda \sin2\beta) \\
  \frac{1}{2}(m^2_Z-\lambda^2 v^2)\sin4\beta & m^2_{\hat H} &
  \lambda v \Lambda \cos2\beta \\
  \lambda v (2\mu-\Lambda \sin2\beta) & \lambda v \Lambda \cos2\beta
  & m^2_{\hat s} \\
\end{array}%
\right),
\eea
the SM-like Higgs boson generally appears as a mixture of CP-even Higgs
bosons:
\bea
h = c_{\theta_1}c_{\theta_2} \hat h
- s_{\theta 1}\hat H - c_{\theta_1}s_{\theta_2} \hat s,
\eea
which we assume to be the resonance near 125 GeV observed at the LHC.
Here $s_\theta= \sin\theta$ and $c_\theta=\cos\theta$, for
$-\pi/2<\theta_i\leq \pi/2$ ($i=1,2,3$) being the mixing angles
in the orthogonal matrix diagonalizing $\hat M^2$ (see eq. (\ref{U-matrix})).
There are also two other mass eigenstates, $s$ and $H$ with
mass $m_s$ and $m_H$, respectively, where we define $H$ to be the
heaviest Higgs boson and fix the mass of $h$ to be $m_h\simeq 125$ GeV.

Let us summarize the properties of the SM-like Higgs boson in the presence
of mixing effects.
The Higgs couplings are determined by
\bea
\label{effective-C}
C_V &=& c_{\theta_1}c_{\theta_2},
\nonumber \\
C_t &=& c_{\theta_1}c_{\theta_2} + s_{\theta_1}\cot\beta,
\nonumber \\
C_b &=& C_\tau = c_{\theta_1}c_{\theta_2} - s_{\theta_1}\tan\beta,
\eea
at the tree-level.
The branching fraction of $h$ for each decay process deviates from the SM value
if the above effective couplings have values different from each other.
This is indeed the case due to the second term in the effective coupling to
the quarks, which relies on $\tan\beta$ and $\theta_1$.
Note that the sign of $\theta_1$ is also important.
For instance, if $\theta_1$ has a positive value, it is possible to reduce
the Higgs coupling to the bottom quark relative to the SM value,
more significantly than those to the $W/Z$ boson and top quark.
This does not happen in the MSSM unless there are large radiative corrections
to the mixing between two doublet Higgs bosons.
Here we only consider $\tan\beta>1$ as required for the top Yukawa coupling
to remain perturbative at high energy scales.
However the situation changes in the NMSSM, where one can consider
$\lambda v > m_Z$ for which $\hat M^2_{12}$ is positive, or sizable mixing
between the singlet and doublets.
As we will see soon, mixing with $\theta_1>0$ has an interesting implication
for the modification in the Higgs signal rate.

On the other hand, mixing effects on the Higgs boson mass can be examined
using the relation,
\bea
\label{mass-relation}
m^2_h = m^2_{\hat h}
- \frac{(s_{\theta_2}s_{\theta_3}-s_{\theta_1}c_{\theta_2}c_{\theta_3})^2}
{c^2_{\theta_1}c^2_{\theta_2}}
(m^2_H-m^2_{\hat h})
- \frac{(s_{\theta_2}c_{\theta_3}+s_{\theta_1}c_{\theta_2}s_{\theta_3})^2}
{c^2_{\theta_1}c^2_{\theta_2}}
(m^2_s-m^2_{\hat h}).
\eea
The above relation imposes a constraint on $m_H$, $m_s$ and the mixing angles
because $m_{\hat h}$ is restricted to be less than a certain value:
\bea
m^2_{\hat h} = m^2_0 + (\lambda^2 v^2 - m^2_Z)\sin^2 2\beta,
\eea
where $m_0$ is equal to $m_Z$ at the tree-level, and can be regarded as
the SM-like Higgs boson mass in the decoupling limit of the MSSM
at large $\tan\beta$.
The NMSSM contribution raises $m_{\hat h}$ above $m_0$ for
$\lambda v > m_Z$, but the amount decreases as $\tan\beta$ increases.
Including radiative corrections, one finds \cite{Okada:1990vk}
\bea
m^2_0 = m^2_Z
+ \frac{3m^4_t}{4\pi^2v^2} \ln\left(\frac{m^2_{\tilde t}}{m^2_t}\right)
+ \frac{3m^4_t}{4\pi^2v^2}\left( X^2_t - \frac{1}{12}X^4_t\right)
+ \cdots,
\eea
where $m_{\tilde t}$ is the stop mass, and $X_t=(A_t-\mu\cot\beta)/m_{\tilde t}$
denotes the stop mixing parameter.
For the stop masses at the TeV scale, $m_0$ is around 115 GeV.
The stop mixing raises $m_0$ further approximately
by the amount, $5\,{\rm GeV}\times(X^2_t-X^4_t/12)$, which is about 5 GeV at $X_t=1$.
The stop mixing contribution is maximized at $X_t=\sqrt 6$, where
$m_0$ can reach about 130 GeV for stop masses around TeV.
In section \ref{sec:constraints}, taking the value of $m_0$ in the range
between 95 and 135 GeV, we will examine in detail the region of parameter
space consistent with various constraints on the Higgs mixing.

\subsection{Mixing effects on the Higgs signal rate}

The observed Higgs signal rate in the $WW/ZZ$ channel, which has a relatively
small experimental error compared to other decay processes,
is in good agreement with the SM predictions.
However this does not necessarily imply small mixing in the neutral Higgs
sector.
It is thus of importance to examine the signal rate in other channels when
the SM-like Higgs boson has $R^{VV}_h\approx 1$ in the presence of Higgs
mixing.
The signal rate for $h\to b\bar b$ and $h\to \gamma\gamma$ are modified
according to the relation (\ref{Rii}).

\begin{figure}[t]
\begin{center}
\begin{minipage}{16.4cm}
\centerline{
{\hspace*{0cm}\epsfig{figure=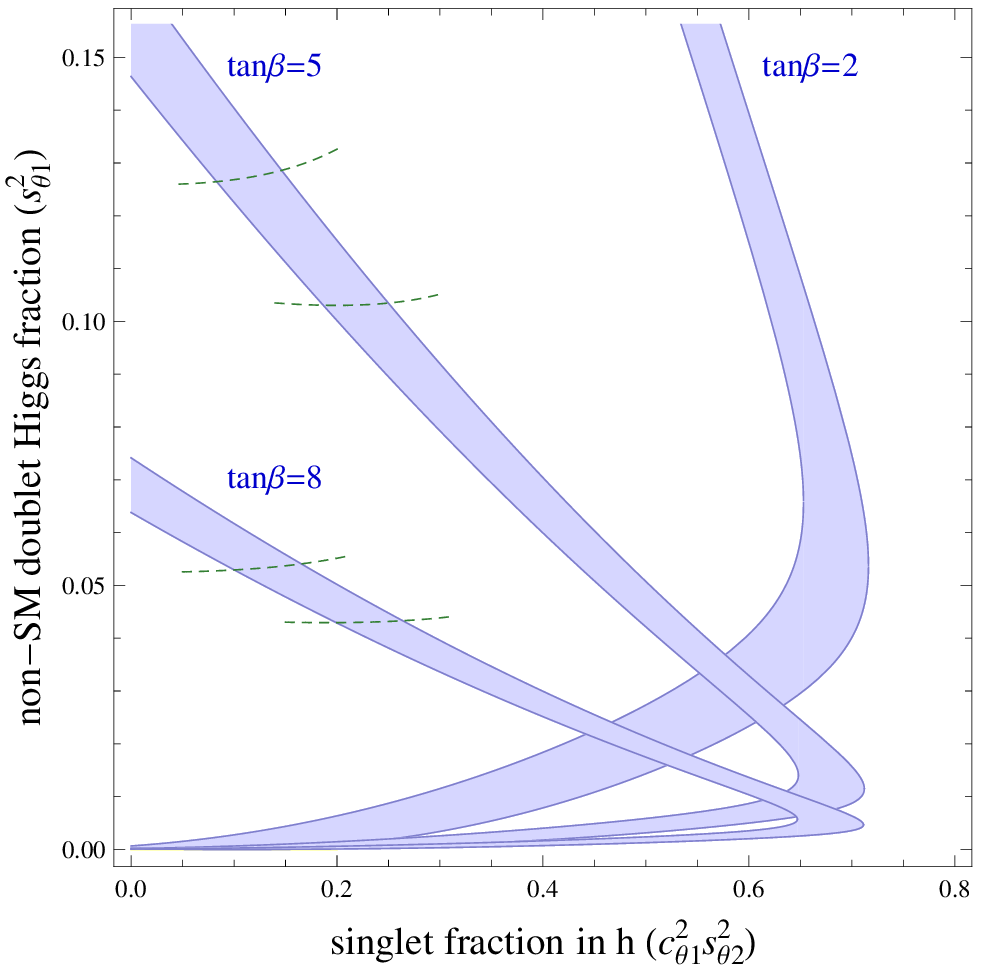,angle=0,width=7.6cm}}
{\hspace*{.6cm}\epsfig{figure=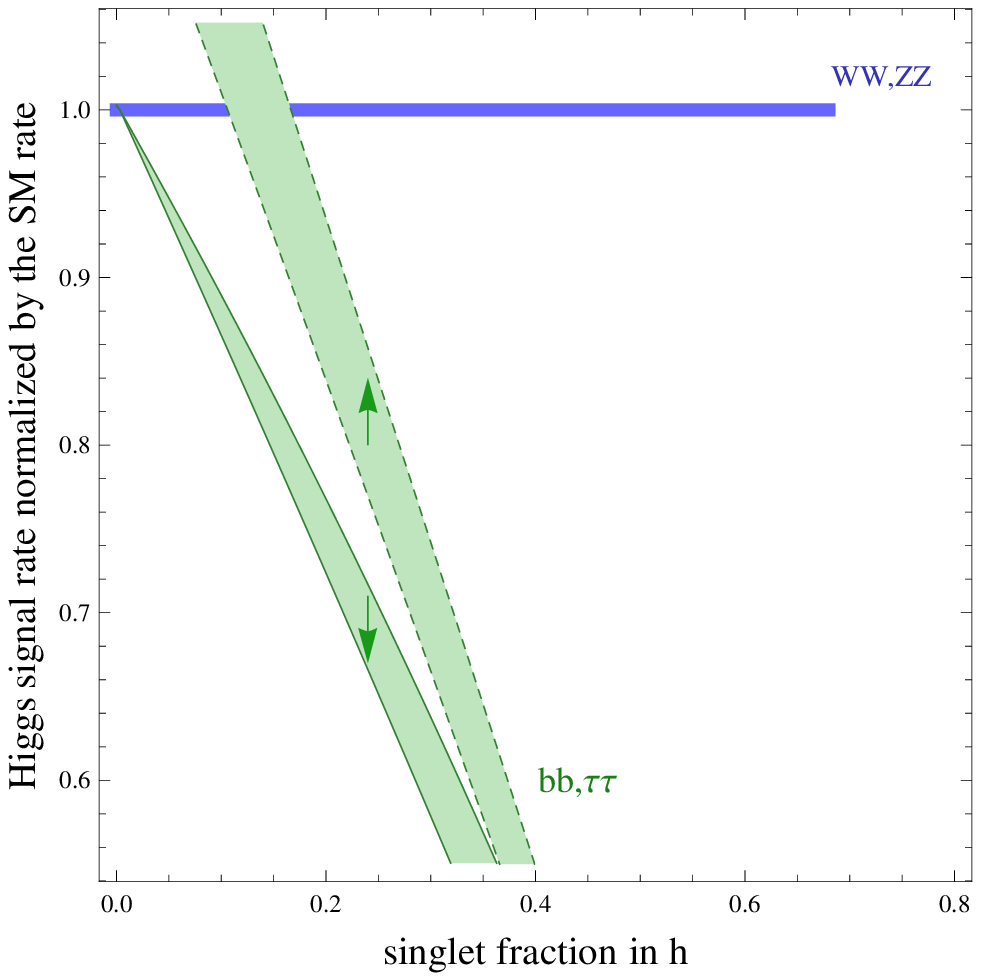,angle=0,width=7.6cm}}
}
\caption{
Mixing effects on the signal rate of the SM-like Higgs boson,
$h=c_{\theta_1}c_{\theta_2}\hat h - s_{\theta_1}\hat H -c_{\theta_1} s_{\theta_2}\hat s$
with mass $m_h=125$ GeV.
The left panel shows the region of parameter space where the signal rate
in the $WW/ZZ$ channel is $R^{VV}_h=1\pm 0.1$, for $\tan\beta=2,5,8$, respectively.
The $b\bar b$ channel also has $R^{b\bar b}_h=1\pm 0.1$ in the region between two dashed
green lines, in the blue band.
The right panel illustrates how the signal rates in $h\to b\bar b$
and $h\to\tau\tau$ are affected by mixing effects when the Higgs boson
has $R^{VV}_h=1$.
The green band is $R^{b\bar b}_h(=R^{\tau\tau}_h)$ obtained
for $2\leq\tan\beta\leq5$, where the colored arrow indicates its behavior
for increasing $\tan\beta$.
The lower (upper) band corresponds to the $b\bar b$ rate for $\theta_1$ fixed at
the smaller (larger) solution of the equation $R^{VV}_h(\theta_1)=1$.
}
\label{fig:Higgs-R-rate}
\end{minipage}
\end{center}
\end{figure}

Plugging the effective Higgs couplings (\ref{effective-C}) into
the relation (\ref{RVV}), we find that Higgs mixing leaves the signal rate
$R^{VV}_h$ close to 1 at two different positive values of $\theta_1$
for given $\tan\beta$ and $\theta_2$.
The larger solution of $R^{VV}_h\approx 1$ is approximately given by
\bea
\label{large-theta1}
\theta_1|_{\rm large} \approx \frac{2.1-1.9 \sin^2\theta_2}{\tan\beta},
\eea
for moderate and large values of $\tan\beta$.
It is also straightforward to see that the smaller solution reads
\bea
\label{small-theta1}
\theta_1|_{\rm small} \approx
\frac{\tan\beta}{1.4\tan^2\beta+1.7}\sin^2\theta_2,
\eea
for $\tan\beta\gtrsim 2$.
Thus $C_V$ and $C_t$ are positive for both solutions, while $C_b$ is found
to be positive only for $\theta_1|_{\rm small}$.
This results in that the $b\bar b$ signal rate can also remain close
to the SM value for $\theta_1|_{\rm large}$ if $C_V\approx -C_b$,
which we find the case in the region around $\sin^2\theta_2=0.15$.
Note also that the gluon-gluon fusion for the Higgs boson production
occurs through the coupling $C_g$, and thus depends on the relative sign
of $C_t$ and $C_b$ since the quarks have masses, $2m_b<m_h<2m_t$.
On the other hand, for the smaller solution $\theta_1|_{\rm small}$,
the SM-like Higgs boson contains non-SM Higgs doublet component whose
fraction is much smaller than the singlet fraction.
In this case, $R^{b\bar b}_h$ is always smaller than $R^{VV}_h$.

Fig. \ref{fig:Higgs-R-rate} shows the region of parameter space leading to
$R^{VV}_h=1\pm0.1$, for $\tan\beta=2,5,8$.
One sees that the Higgs signal rate in the $WW/ZZ$ channel can remain consistent
with the observation even if the SM-like Higgs boson has a large singlet
fraction.
In the right panel of the figure, we illustrate how the $b\bar b$ and $\tau\tau$
rates are modified when the Higgs mixing leads to $R^{VV}_h=1$.
The green band gives $R^{b\bar b}_h(=R^{\tau\tau}_h)$ for $2\leq \tan\beta\leq 5$,
where the colored arrow shows how the signal rate changes as $\tan\beta$ increases.
The lower (upper) green band corresponds to $R^{b\bar b}_h$ for $\theta_1$ fixed
at the smaller (larger) solution of the equation $R^{VV}_h(\theta_1)=1$.
The $b\bar b$ and $\tau\tau$ rates are reduced relative to the SM case for
the smaller solution, however this is not necessarily the case for the other
solution.
On the other hand, the mixing also affects the diphoton rate, which will be
examined in section \ref{sec:higgsino-effect} by including the contribution
from the charged higgsino.

\section{Enhanced diphoton rate by higgsino effects}
\label{sec:higgsino-effect}

The Higgs signal in the $h\to \gamma\gamma$ process has been observed at the
rate, 1.5 to 1.8 times larger than the value predicted in the SM.
Although the statistics is small yet compared to the $WW/ZZ$ channel, this may
be a hint for physics beyond the SM.
In particular, it is interesting to see if the observed enhancement of the
diphoton rate can be explained within the NMSSM, where the deviation
in the $WW/ZZ$ channel is small even if the Higgs mixing is sizable.
Indeed the Higgs coupling to photons can receive a significant contribution
from the charged-higgsino loop when the Higgs boson has a large singlet
component.

Let us examine the higgsino effects on the diphoton rate in the presence
of Higgs mixing.
Since the singlet scalar couples to the higgsinos through the Yukawa term
$\lambda S \tilde H_u \tilde H_d$ in the NMSSM,
the SM-like Higgs boson has the interaction,
\bea
{\cal L} = \frac{\lambda c_{\theta_1}s_{\theta_2}}{\sqrt2}
h \tilde H^+_u \tilde H^-_d + \cdots.
\eea
Thus the Higgs coupling to photons is radiatively generated by the charged-higgsino
loop.
The contribution to $C_\gamma$ is found to be \cite{Carmi:2012in},
\bea
\delta C_\gamma|_{\rm SUSY} \simeq
-0.17 A_f(\tau_{\tilde H^\pm})
\frac{\lambda v }{m_{\tilde H^\pm}}\,c_{\theta_1}s_{\theta_2},
\eea
where $\tau_i = m^2_h/4m^2_i$, and the charged higgsino has mass
$m_{\tilde H^\pm}\simeq |\mu|$
simply assuming that the Wino is heavy so that the mixing in the chargino
sector is small.
The loop function is given by
$A_f(\tau)=3(\tau+(\tau-1)\arcsin^2\sqrt\tau)/2\tau^2$ for $\tau\leq 1$, and
is approximated as $A_f(\tau)\simeq 1+7\tau/30$ for small values of $\tau$.
On the other hand, the effective couplings $C_V$, $C_t$ and $C_b$
are negligibly affected by the charged higgsino.

\begin{figure}[t]
\begin{center}
\begin{minipage}{16.4cm}
\centerline{
{\hspace*{0cm}\epsfig{figure=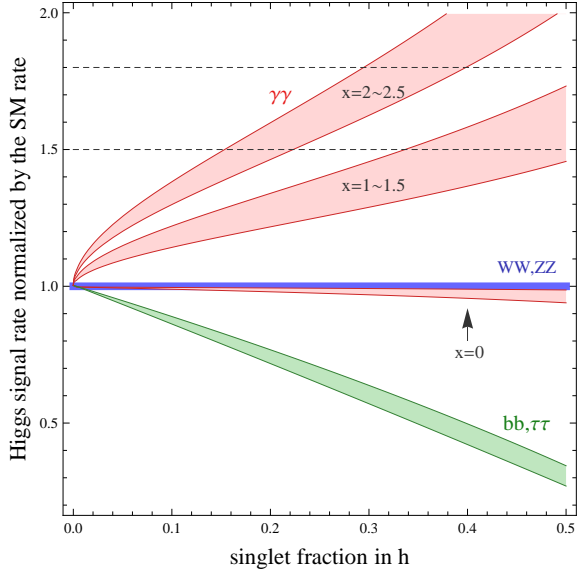,angle=0,width=7.6cm}}
}
\caption{
Enhanced diphoton rate via the charged-higgsino loop contribution, in the presence
of Higgs mixing with positive $\theta_1$ and $\theta_2$.
We fix the value of $\theta_1$ at the smaller solution of $R^{VV}_h(\theta_1)=1$,
for which the diphoton rate is determined by $\theta_2$ and
$x\equiv \lambda v/|\mu|$, almost independent of $\tan\beta$.
The red band corresponds to $R^{\gamma\gamma}_h$ for $x=0$, $1\leq x\leq 1.5$,
and $2\leq x\leq 2.5$ from below, respectively.
Here we take $2\leq \tan\beta\leq 8$.
The figure also shows the signal rate in other channels, which are rarely
affected by the charged-higgsino loop.
The green band gives $R^{b\bar b}_h$ obtained for $\tan\beta$ in the same range.
}
\label{fig:Diphoton}
\end{minipage}
\end{center}
\end{figure}

The SM-like Higgs boson can have $R^{VV}_h\approx1$ for the Higgs mixing
(\ref{large-theta1}) or (\ref{small-theta1}), for which the diphoton signal
rate will be reduced a bit below the SM value if $\delta C_\gamma|_{\rm SUSY}=0$.
However, including the radiative correction from $\tilde H^\pm$,
the diphoton rate reads
\bea
\label{R-rr-higgsino}
R^{\gamma\gamma}_h
\simeq \frac{1.52 C^2_\gamma}{C^2_V}R^{VV}_h
\simeq
\left( 1
- 0.28\frac{s_{\theta_1}}{c_{\theta_1}c_{\theta_2}}\frac{1}{\tan\beta}
+ 0.21\frac{s_{\theta_2}}{c_{\theta_2}} \frac{\lambda v}{|\mu|}
\right)^2 R^{VV}_h,
\eea
for $|\mu|\gtrsim 100$ GeV,
and thus an enhancement of diphoton signals relative to the SM case
is achievable for $R^{VV}_h\approx1$, when $\theta_2$ is positive.
Note that, if $\delta C_\gamma|_{\rm SUSY}=0$, the Higgs signal rate in
each channel is modified by the mixing effects independently of the sign
of $\theta_2$.
To get a large enhancement, the charged higgsino needs to have a small mass.
In addition, we need a large value of $\theta_2$, which determines the singlet
fraction of the SM-like Higgs boson.
Since Higgs mixing with positive $\theta_1$ allows $h$ to contain a large
singlet component while having $R^{VV}_h\approx 1$,
the second term in the bracket in (\ref{R-rr-higgsino}) tells that
a small value of $\theta_1$ is favored to have $R^{\gamma\gamma}_h$ larger
than $R^{VV}_h$.
For the smaller solution (\ref{small-theta1}) of $R^{VV}_h(\theta_1)\approx 1$,
the diphoton rate becomes insensitive to $\tan\beta$, and is written
\bea
R^{\gamma\gamma}_h \approx
\left(1
+ 0.21 \frac{s_{\theta_2}}{c_{\theta_2}}\frac{\lambda v}{|\mu|}\right)^2.
\eea
Thus the signal rate in $h\to \gamma\gamma$ can be larger than the SM expectation,
but still with little deviation in the $WW/ZZ$ channel.
The enhancement crucially depends on the value of $\lambda v/|\mu|$,
which is less than about 3.5 for $\lambda<2$ and $\mu$ is larger than
100 GeV.
In section \ref{sec:constraints}, we will examine the value of
$\lambda v/|\mu|$ consistent with various constraints.

Fig. \ref{fig:Diphoton} illustrates the effect of the charged higgsino
on the diphoton rate, where we have fixed the value of $\theta_1$
to give $R^{VV}_h=1$ for positive $\theta_1$ and $\theta_2$.
Among the two solutions of $R^{VV}_h(\theta_1)=1$, we choose the smaller one,
for which the diphoton rate is almost independent of $\tan\beta$.
The $R^{\gamma\gamma}_h$ is shown in the red band for $x=0$,
$1\leq x\leq 1.5$ and $2\leq x \leq 2.5$, where $x\equiv\lambda v/|\mu|$.
In the absence of the charged-higgsino contribution, which is the case
with $x=0$, the diphoton rate is slightly below the SM value.
However the higgsino effect can significantly enhance diphoton signals,
which demands a large $x$ and sizable singlet fraction in $h$.
For instance, the diphoton rate is about 1.5 times larger than the SM value
at $x=1.2$ if the singlet fraction in $h$ is 0.4.
On the other hand, the green band indicates $R^{b\bar b}_h$ obtained
for $2\leq \tan\beta \leq 8$.
The $b\bar b$ rate is reduced relative to the SM value.
Meanwhile, one may consider Higgs mixing such that $R^{VV}_h$ is
a bit larger than 1.
Then the diphoton rate will be increased by the similar amount,
whereas $R^{b\bar b}_h$ will decrease further.

\section{Constraints on Higgs mixing}
\label{sec:constraints}

In this section, we discuss various constraints on the mass spectrum of
Higgs bosons and the mixing angles which arise because of their relations
to the model parameters in $\hat M^2$.
Then we will consider several examples to explore the region of parameter
space where the SM-like Higgs boson has $R^{VV}_h\approx1$ while satisfying
the constraints.
The $WW/ZZ$ rate remains close to the SM value at the two solutions of
$R^{VV}_h(\theta_1)\approx1$, among which the smaller one is favored to get
a large enhancement in the diphoton rate through the charged-higgsino
contribution.
So we will focus on the smaller solution, and examine the diphoton rate
in the explicit examples.

\subsection{Relation with the Lagrangian parameters}

Let us first discuss the constraints on Higgs mixing related to the model
parameter $\mu$ and $\lambda$, which determine the off-diagonal components
of $\hat M^2$.
Their values are determined once we fix the mass eigenvalues and mixing angles.
This implies that $\mu$ and $\lambda$ are written as functions of
\bea
\label{mixing-parameters}
\xi \equiv (\tan\beta,m_H,m_s,\theta_1,\theta_2,\theta_3).
\eea
Meanwhile the LEP bound on the chargino mass requires $\mu$ larger
than about 100 GeV \cite{PDG}.
In addition, in order for the model to remain perturbative up to
at least 10 TeV, $\lambda$ should be less than about 2 at the weak scale.
For NMSSM models with $\lambda$ larger than about 0.7 \cite{Barbieri:2006bg},
there should appear new physics, such as strong dynamics involving $S$,
at an intermediate scale below the conventional GUT scale $\sim 10^{16}$ GeV.
The above bounds on $\mu$ and $\lambda$ translate into the constraints
on $\xi$.

Here we present the relation between $\xi$ and the model parameters
appearing in the off-diagonal elements of $\hat M^2$:
\bea
\lambda^2 v^2 &=& m^2_Z
+ \frac{1}{\sin4\beta}
\Big( (m^2_H - m^2_s) s_{\theta_2} s_{2\theta_3}
+ 2(m^2_h - m^2_H c^2_{\theta_3} - m^2_s s^2_{\theta_3})
s_{\theta_1}c_{\theta_2}
\Big) c_{\theta_1},
\nonumber \\
\lambda v \mu &=&
-\frac{1}{4} m^2_h c^2_{\theta_1} s_{2\theta_2}
-\frac{1}{4}(m^2_H-m^2_s) s_{\theta_1} c_{2\theta_2}s_{2\theta_3}
\nonumber \\
&& +\, \frac{1}{4} \Big(
(m^2_H-m^2_s s^2_{\theta_1}) s^2_{\theta_3}
- (m^2_H s^2_{\theta_1} - m^2_s) c^2_{\theta_3}
\Big)s_{2\theta_2}
\nonumber \\
&& -\, \frac{\tan2\beta}{4} \Big(
(m^2_H-m^2_s) c_{\theta_2} s_{2\theta_3}
 - 2(m^2_h-m^2_H c^2_{\theta_3} -m^2_s s^2_{\theta_3})
s_{\theta_1}s_{\theta_2}
\Big)c_{\theta_1},
\nonumber \\
\lambda v \Lambda &=& -\frac{1}{2\cos2\beta}
\Big( (m^2_H - m^2_s) c_{\theta_2} s_{2\theta_3}
- 2(m^2_h - m^2_H c^2_{\theta_3} - m^2_s s^2_{\theta_3})
s_{\theta_1}s_{\theta_2}
\Big)c_{\theta_1},
\eea
where we have used that the orthogonal matrix $U$ which diagonalizes
$\hat M^2$ is parameterized in terms of three mixing angles,
\bea
\label{U-matrix}
U = \left(%
\begin{array}{ccc}
  c_{\theta_1}c_{\theta_2} & -s_{\theta_1} & -c_{\theta_1}s_{\theta_2} \\
  s_{\theta_1}c_{\theta_2}c_{\theta_3}
  -s_{\theta_2}s_{\theta_3} & c_{\theta_1}c_{\theta_3} &
  -c_{\theta_2}s_{\theta_3} -s_{\theta_1}s_{\theta_2}c_{\theta_3}  \\
  s_{\theta_1}c_{\theta_2}s_{\theta_3} + s_{\theta_2}c_{\theta_3} &
  c_{\theta_1}s_{\theta_3} &
  c_{\theta_2}c_{\theta_3}-s_{\theta_1}s_{\theta_2}s_{\theta_3} \\
\end{array}%
\right),
\eea
for $c_\theta=\cos\theta$ and $s_\theta=\sin\theta$.

\subsection{Higgs boson mass}

There are also constraints related to the mass spectrum of the Higgs
bosons.
The SM-like Higgs boson has a mass which is written as a function
of $m_0$ and $\xi$, as is given by (\ref{mass-relation}).
Thus, requiring $m_h\simeq 125$ GeV constrains the parameter space of
$m_0$ and $\xi$.
Here the value of $m_0$ has an upper bound, crucially depending on the
radiative corrections.
For instance, $m_0$ is less than about 120 GeV for the stop with
mass around or smaller than 1 TeV and small stop mixing.
One can also arrange $m_0$ around or larger than 130 GeV by considering
stop masses above a few TeV or invoking large stop mixing.
It is worth noting that, in the NMSSM, $m_0$ can receive an additional sizable
contribution from the higgsino loops.
The superpotential $f(S)$ provides mass to the singlino,
$m_{\tilde S} = \langle \partial^2_S f \rangle$.
If $m_{\tilde S}$ is around or below $\mu$, the interactions of
the higgsinos ($\tilde H_u$ and $\tilde H_d$) and the singlino ($\tilde S$),
\bea
-{\cal L}_{\rm int} =
\lambda H_u \tilde H_d \tilde S
+ \lambda H_d \tilde H_u \tilde S
+ \mu \tilde H_u \tilde H_d
+ \frac{1}{2}m_{\tilde S} \tilde S \tilde S
+ {\rm h.c.},
\eea
will remain relevant at low energy scales.
Then the mass of $\hat h$ receives additional radiative corrections,
analogously to that from the top Yukawa coupling.
The correction is estimated naively to be \cite{PQ-NMSSM}
\bea
\delta m^2_0 \approx \frac{\lambda^4 v^2}{4\pi^2}\ln\left(
\frac{m^2_s}{|\mu|^2}\right),
\eea
for $m^2_s\gg |\mu|^2$.
The above contribution is quite sensitive to $\lambda$, but is almost
independent of $\tan\beta$.
In the analysis, we will consider $95\,{\rm GeV}\leq m_0\leq 135\,{\rm GeV}$,
while keeping in mind that larger values can be allowed depending on models.

On the other hand, the experimental constraints from the $b\to s\gamma$ process
require the charged Higgs scalar to be heavier than about 350 GeV
\cite{Gambino:2001ew}
unless there are cancellations with superparticle contributions.
The charged Higgs scalar obtains a mass according to
$m^2_{H^\pm}=m^2_{\hat H}+m^2_W-m^2_Z\sin^22\beta$ at the tree-level,
while $m^2_{\hat H}$ is given by
\bea
m^2_{\hat H} = s^2_{\theta_1} m^2_h + c^2_{\theta_1}c^2_{\theta_3}m^2_H
+c^2_{\theta_1}s^2_{\theta_3}m^2_s,
\eea
implying that an addition constraint will be imposed if one requires
a large value of $m_{H^\pm}$ to avoid the constraint from $b\to s\gamma$.

For a large value of $m_H$, the mixing angle $\theta_3$ needs to be small
since otherwise mixing between $\hat h$ and $\hat H$ will decrease the
mass of the SM-like Higgs boson significantly.
Suppose that $\theta_3$ is similar to or smaller than $\theta_1$ in size.
Then the Higgs mass reads
\bea
m^2_h \approx m^2_{\hat h}
- \frac{1}{\tan^2\beta} \frac{s^4_{\theta_2}}{1-s^2_{\theta_2}}
(m^2_H-m^2_{\hat h})
- \frac{s^2_{\theta_2}}{1-s^2_{\theta_2}}
(m^2_s-m^2_{\hat h}),
\eea
for $\theta_1=\theta_1|_{\rm small}$, where the coefficients of order unity
have been neglected in the last two terms.
We see that the second term, which is a result of mixing between two doublet
Higgs bosons, is much suppressed compared to the third term.
Indeed one can find that the second term lowers the mass of $h$ by less than
about 10 GeV even for $H$ having mass around or above 350 GeV, if
\bea
s^2_{\theta_2}
\lesssim 0.15\times\left(\frac{m_H}{350{\rm GeV}}\right)^{-1}\tan\beta.
\eea
for $\tan\beta\gtrsim 2$.
On the other hand, the third term arises due to mixing with the singlet,
and can have either sign.
This contribution will compensate the mass decrease by the second term if
$s$ has a mass smaller than $m_{\hat h}$.
The relation $m^2_{\hat h}=m^2_0 + (\lambda^2v^2-m^2_Z)\sin^22\beta$ tells
that the sign of the third term depends $m_0$, $\lambda$ and $\tan\beta$.
It is clear that $m_s$ much larger than $m_0$ will make it difficult
to get $m_h\simeq 125$ GeV unless the NMSSM contribution to $m_{\hat h}$
is large enough.

Let us discuss further the properties of $s$, which is singlet-like
in most of the parameter space of interest.
Since it couples to the SM particles through the doublet Higgs components,
it may be possible to detect $s$ at hadron colliders when Higgs mixing
is large.
The production and decay rate of $s$ depend on its mass and the fraction
of the Higgs doublets.
However, if the invisible decay into neutralinos is kinematically allowed,
the discovery of the singlet-like Higgs boson will become very difficult.
This is the case for $m_h<2m_{\chi^0_1}<m_s$, where the lower bound on
$m_{\chi^0_1}$ would be necessary to suppress the invisible decay
of the SM-like Higgs boson.\footnote{
In models where $f(S)$ contains only a linear term of $S$
\cite{PQ-NMSSM,Panagiotakopoulos:1999ah,Panagiotakopoulos:2000wp,Bae:2012am},
the singlino has no Majonara mass term, and the mass and the coupling
to $h$ of $\chi^0_1$ are proportional to $\lambda^2 v\sin2\beta/\mu$.
One would thus need small $\tan\beta$ and large $\lambda$ to kinematically
forbid the invisible decay of $h$ into neutralinos.
}
On the other hand, if $m_s<m_h$, one should take into account
the LEP constraints on the signal rate of $s$ in $e^+e^-\to Zs \to Zb\bar b$
\cite{LEP-Higgs}.
In this case, it may still be possible to have sizable mixing if
$m_s$ lies in the range between about 90 and 110 GeV.

Before proceeding to the analysis, we briefly mention about the property
of the lightest neutralino, which we assume the lightest superparticle
(LSP).
For $\lambda v/|\mu|$ around one, the lightest neutralino $\chi^0_1$ is a
sizable mixture of the neutral higgsinos and singlino, simply assuming
that all the gauginos are heavy.
If $\chi^0_1$ is heavier than the $W$-boson, the annihilation among
LSPs takes place with a large cross section, dominantly through
the $t$-channel charged-higgsino exchange with $W$ bosons in the final state.
Then thermal relic density of $\chi^0_1$ would be too small to account
for the observed dark matter density.
To avoid this, one may consider light gauginos so that
$\chi^0_1$ has a sizable Bino component.
Another way would be to implement non-thermal production of LSPs, or
to consider the QCD axion as a dark matter.

On the other hand, for $m_{\tilde s}\gg|\mu|$ and $\lambda v/|\mu|$
around one, the mixing decreases the mass of a higgsino-like neutralino, and
thus it is possible to have $\chi^0_1$ lighter than $m_W$ but heavier
than $m_h/2$.
In such a case, the annihilation via the $s$-channel
$Z$-boson exchange into SM fermions is not so effective due to the
suppressed $Z \chi^0_1 \chi^0_1$ coupling, and the coannihilation
effect \cite{Mizuta:1992qp} is also suppressed because
the charged-higgsino still has a mass equal to $\mu$.
However, since the $h \chi^0_1 \chi^0_1$ coupling can be sizable
in the NMSSM, it would be possible to get the correct dark matter
density from thermally produced LSPs.
A detailed analysis will be given elsewhere.

\subsection{Examples}

We have discussed the constraints and the favored values of model parameters.
Let us now examine the diphoton rate in the region of parameter space where
such requirements are satisfied.
In the analysis, we impose the constraints,
\bea
m_h \simeq 125\,{\rm GeV},\quad
|\mu| \geq 105\,{\rm GeV},\quad
\lambda \leq 2,
\eea
by taking $m_0$ in the range
\bea
95\,{\rm GeV} \leq m_0 \leq 135\,{\rm GeV}.
\eea
For given $m_H$, $m_s$ and $\theta_3$, three parameters of $\xi$ in
(\ref{mixing-parameters}) remain unfixed.
Among them, $\theta_1$ is fixed if one requires the signal rate in
the $WW/ZZ$ channel to be close to the SM value.
Then we can examine the above constraints by scanning over $\tan\beta$
and $\theta_2$, or equivalently in terms of $\tan\beta$ and the singlet
fraction in the SM-like Higgs boson.
The Higgs signal rate in $h\to\gamma\gamma$ is determined
by $x=\lambda v/|\mu|$ and the singlet fraction in $h$.

\begin{figure}[t]
\begin{center}
\begin{minipage}{16.4cm}
\centerline{
{\hspace*{0cm}\epsfig{figure=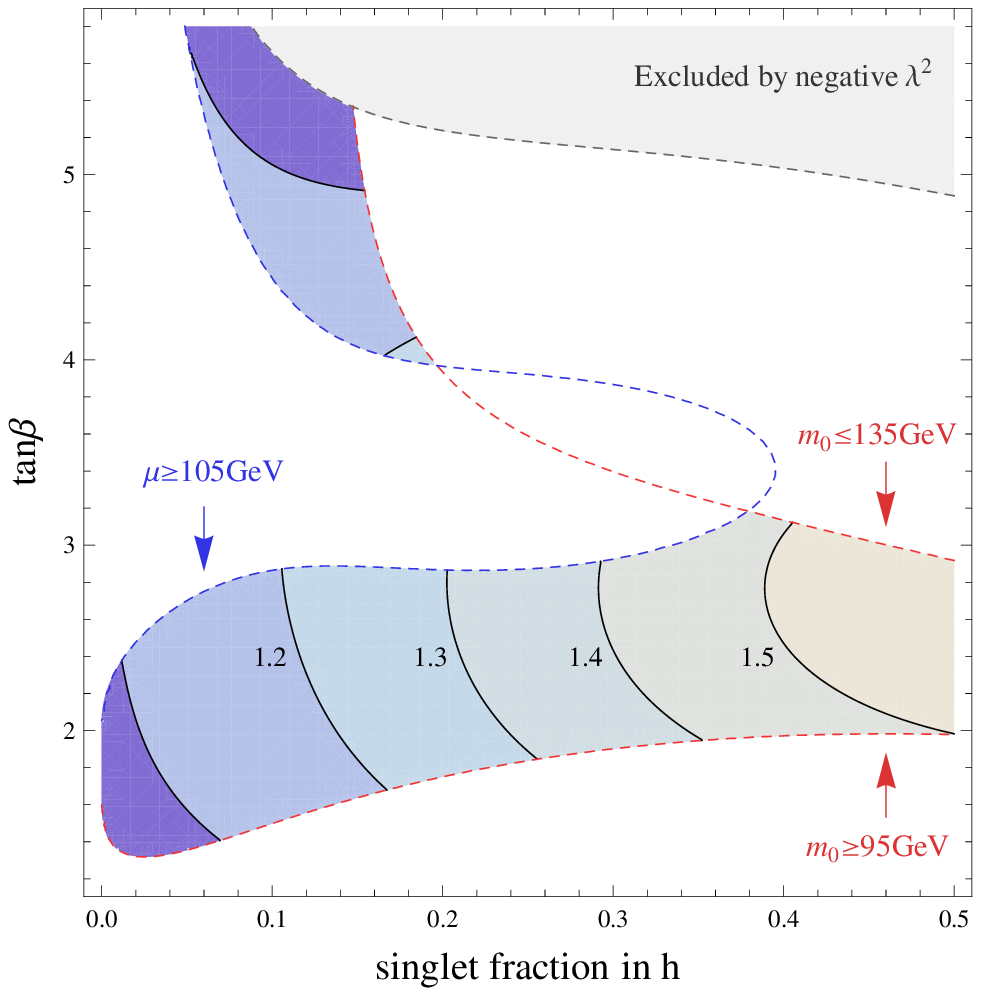,angle=0,width=7.6cm}}
{\hspace*{.6cm}\epsfig{figure=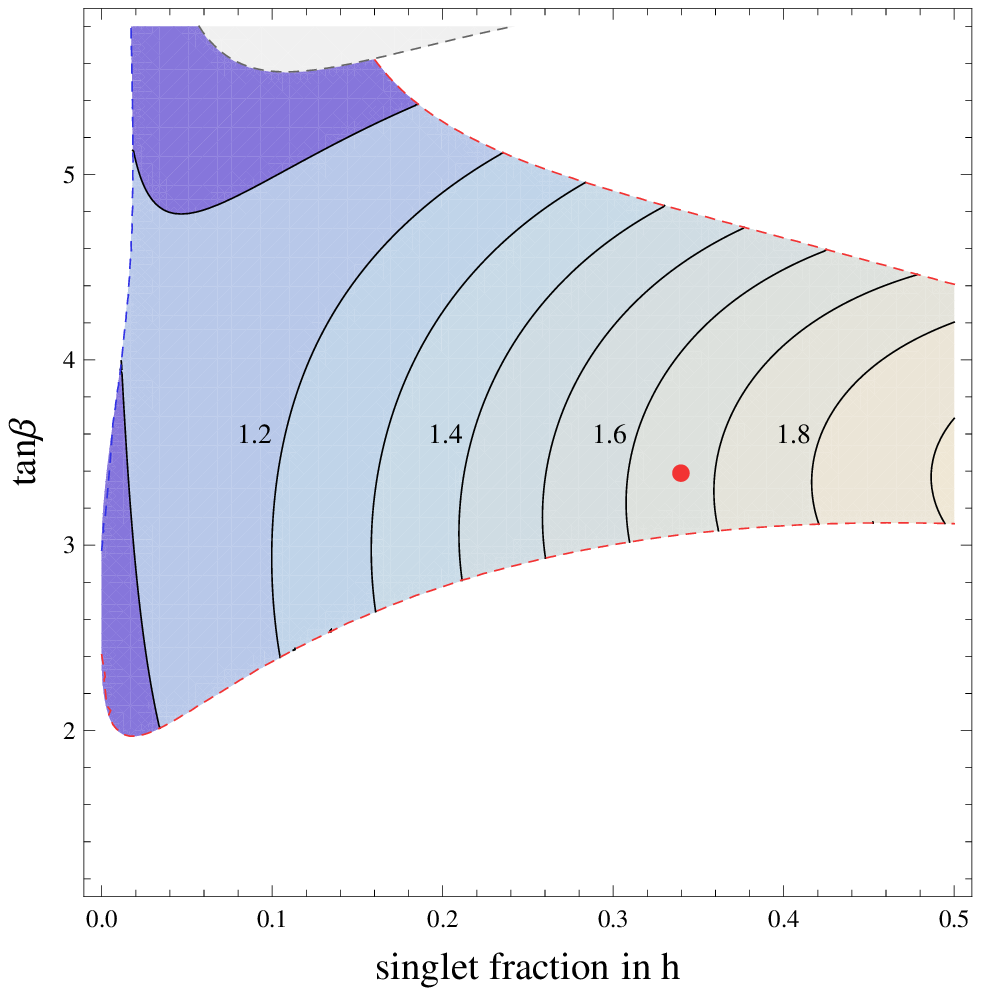,angle=0,width=7.6cm}}
}
\caption{
Enhanced diphoton rate in the region of parameter space allowed by
$|\mu|\geq 105$ GeV and $\lambda\leq2$,
where we have fixed $\theta_1$ at the smaller solution of $R^{VV}_h(\theta_1)=1.05$.
The left panel is for the case with
$(m_H,m_s,\theta_3)=(480{\rm GeV},160{\rm GeV},0.1)$,
while the right one is for $(m_H,m_s,\theta_3)=(780{\rm GeV},160{\rm GeV},0.07)$.
In both figures, $\mu$ is larger than 105 GeV in the right side of the
blue-dashed curve, and $m_h=125$ GeV is obtained for
$95\,{\rm GeV}\leq m_0\leq 135\,{\rm GeV}$ in the region between two red-dashed
curves.
The gray-shaded region is excluded due to negative $\lambda^2$.
The contours of $R^{\gamma\gamma}_h$ are indicated by black lines, with
the contour interval equal to 0.1.
To see the detailed Higgs properties, we take an example point as given
by the red-filled circle.
}
\label{fig:examples}
\end{minipage}
\end{center}
\end{figure}

In Fig. \ref{fig:examples}, we show the diphoton rate relative to the SM value
in the region of parameter space allowed by the requirements mentioned above,
for two examples,
\bea
\mbox{\it example 1 }: &&
(m_H,m_s,\theta_3) = (480{\rm GeV},160{\rm GeV},0.10),
\nonumber \\
\mbox{\it example 2 }: &&
(m_H,m_s,\theta_3) = (780{\rm GeV},160{\rm GeV},0.07),
\nonumber
\eea
with $\theta_1$ fixed at the smaller solution of $R^{VV}_h(\theta_1)=1.05$.
For both examples, the parameter space is mainly constrained by the requirements,
$m_h=125$ GeV and $|\mu|\geq105$ GeV.
The right side of the blue-dashed curve gives $\mu$ larger than 105 GeV,
and $h$ has $m_h=125$ GeV between two red-dashed curves if one considers
$95\,{\rm GeV}\leq m_0\leq 135\,{\rm GeV}$.
The gray-shaded region is excluded since $\lambda^2<0$.
The diphoton rate relative to the SM value is shown in black lines.
We see that an enhancement by a factor of 1.5 or larger can be accommodated
through the charged-higgsino contribution, when $\tan\beta$ is about $2-3$
and the singlet fraction in $h$ is larger than 0.39 in the left panel.
For the other example, this is achieved at $3\lesssim \tan\beta\lesssim 4.8$
when the singlet fraction in $h$ is larger than 0.26.
On the other hand, the mixing effects reduce the $b\bar b$ rate
below the SM value.

To see the relation between $\xi$ and the model parameters, we take
an example point at
\bea
(\tan\beta,m_H,m_s,\theta_1,\theta_2,\theta_3) =
(3.4,\,780\,{\rm GeV},\,160\,{\rm GeV},\,0.07,\,0.62,\,0.07),
\eea
which corresponds to the red-filled circle in the figure.
The above set of parameters translates into
\bea
(\lambda,\mu,\Lambda,m_0) \simeq
(1.03,112\,{\rm GeV},376\,{\rm GeV},110\,{\rm GeV}),
\eea
together with $m_{\hat H}\simeq 776$ GeV and $m_{\hat s}\simeq 167$ GeV.
The Higgs signal rate in the $WW/ZZ$ channel is $R^{VV}_h=1.05$
at the example point, and the $b\bar b/\tau\tau$ channel has
$R^{b\bar b}_h=R^{\tau\tau}_h\simeq 0.51$.
Finally, the diphoton rate is found to be
$R^{\gamma\gamma}_h\simeq 1.66$, as a result of the charged-higgsino
loop contribution combined with the Higgs mixing effects.

\section{Conclusions}
\label{sec:conclusion}

We have examined how to arrange a SM-like Higgs boson in the NMSSM,
in a way consistent with the LHC results on the Higgs boson search.
The observed $WW/ZZ$ signal rate can be compatible with the model
even when the SM-like Higgs boson has a large singlet component.
This leads to an interesting possibility that the Higgs to diphoton
signal rate is enhanced by the loops of charged higgsino which couples
to the SM-like Higgs boson through mixing with the singlet scalar.
A large enhancement is achievable when there is a large singlet
fraction in the SM-like Higgs boson and the charged higgsino is light.

The Higgs mixing depends on the higgsino mass parameter $\mu$ and
the singlet coupling $\lambda$ to the Higgs doublets.
Hence there are constraints on the Higgs mixing angles coming
from the LEP bound on the chargino mass, and the perturbativity
limit on $\lambda$.
In addition, the model should provide a mass near 125 GeV to the SM-like
Higgs boson.
Taking into account these constraints, we find that the charged-higgsino
loop contribution, combined with the mixing effects, can enhance the diphoton
signal rate above the SM expectation by a factor of 1.5 or more,
while having little deviation in the $WW/ZZ$ rate.

On the other hand, the Higgs mixing effects reduce the $b\bar b$ and
$\tau\tau$ signals below the SM values, if the mixing between two doublet
Higgs bosons $\hat h$ and $\hat H$ is relatively small, which is favored
to get a sizable enhancement of the diphoton signal rate while keeping
the $WW/ZZ$ rate similar to the SM prediction.
Indeed we could see that the $b\bar b$ and $\tau\tau$ rates can be reduced
by a factor of 0.5, or even less, for the mixing parameters which would enhance
the diphoton signal by a factor of 1.5.

\section*{Acknowledgment}

KC and SHI were supported by the National Research Foundation
of Korea (NRF) grant (No. 2007-0093865 and No. 2012R1A2A2A05003214)
and the BK21 project funded by the Korean Government (MEST).
KSJ and MY were supported by Grants-in-Aid for Scientific Research
from the Ministry of Education, Science, Sports, and Culture (MEXT),
Japan, No. 23104008 and No. 23540283.
\\

\end{document}